\documentclass[preprint,showpacs,preprintnumbers,amsmath,amssymb]{revtex4}

\usepackage{graphicx,subfigure}
\usepackage{dcolumn}
\usepackage{bm}
\usepackage{psfrag}
\usepackage{feynmf}

\newcommand{\be}{\begin{equation}}
\newcommand{\ee}{\end{equation}}
\newcommand{\bea}{\begin{eqnarray}}
\newcommand{\eea}{\end{eqnarray}}

\newcommand{\E}[1]{\;10^{#1}\;}
\newcommand{\e}[1]{\;10^{-#1\;}}
\newcommand{\f}[2]{\frac{#1}{#2}}
\newcommand{\lain}{\cline{2-6}}
\newcommand{\go}{\rightarrow}
\newcommand{\nus}[2]{\bar{\nu}_{#1} \nu_{#2}}
\newcommand{\nee}{\nu_e}
\newcommand{\ane}{\bar{\nu}_e}
\newcommand{\nm}{\nu_\mu}
\newcommand{\anm}{\bar{\nu}_\mu}
\newcommand{\nt}{\nu_\tau}
\newcommand{\ant}{\bar{\nu}_\tau}
\newcommand{\eee}{e^-}
\newcommand{\m}{\mu^-}
\newcommand{\ttt}{\tau^-}
\newcommand{\aee}{e^+}
\newcommand{\am}{\mu^+}

\newcommand{\sg}[1]{\tilde{\sigma}(#1)}
\newcommand{\br}[1]{\tilde{B}(#1)}

\begin{document}

\bibliographystyle{apsrev}

\preprint{CERN-PH-TH/2004-196} \preprint{UAB-FT-574}

\title{Systematic Approach to Gauge-Invariant Relations between Lepton
  Flavor Violating Processes} 

\author{Alejandro Ibarra$^1$}
\author{Eduard Mass{\'o}$^2$}
\author{Javier Redondo$^2$}
\affiliation{$^1$Department of Physics, Theory Division, CERN\\CH-1211
  Geneva 23, Switzerland\\ }
\affiliation{$^2$Grup de F{\'\i}sica Te{\`o}rica and Institut
de F{\'\i}sica d'Altes
Energies\\Universitat Aut{\`o}noma de Barcelona\\
08193 Bellaterra, Barcelona, Spain}


\date{\today}

\begin{abstract}
We analyze four-lepton contact interactions that lead to lepton flavor
violating processes, with violation of individual family lepton
number but total lepton number conserved. 
In an effective Lagrangian framework, the assumption of gauge
invariance leads to relations among branching ratios and cross
sections of lepton flavor violating processes. In this paper, we work
out how to use these relations systematically. We also study the
consequences of loop-induced processes.  

\end{abstract}

\maketitle


\section{Introduction}

The Standard Model (SM) of Particle Physics does not allow
conversion between lepton flavors and thus is not able to
accommodate neutrino oscillation phenomena. Therefore, the
evidence of oscillations (see \cite{Gonzalez-Garcia:2002dz} for a
recent review) makes leptons a promising sector where to look for
clues about new physics. This strongly motivates the quest for
other lepton flavor violating (LFV) searches; a quest that is
already being carried out and that will be pushed in the near
future by a variety of experiments, including the LHC and the
projected neutrino factory or the linear collider.

From a theoretical point of view, LFV originates at high energies
and can be conveniently described in a model independent way
using an effective Lagrangian approach. We will
construct our effective theory using just SM fields. Under this
assumption, the minimal extension of the SM
that can accommodate neutrino masses consists on
adding the following dimension five operator to the
SM Lagrangian \cite{Weinberg:1979sa}:
\begin{equation}
{\cal L}={\cal L}_{SM} + \frac{{\alpha}_{ij}}{\Lambda} (L_i H)^T (L_j H) +h.c.
\label{five}
\end{equation}
This operator violates total
lepton number, $L=\Sigma L_i = L_e+L_\mu+L_\tau$, by two units, and gives rise
to Majorana masses for the neutrinos after the breaking of the
electroweak symmetry. This dimension five operator also violates
flavor, and induces rates for LFV processes suppressed by the scale
$\Lambda$. In view of the measured neutrino mass splittings,
this scale is presumably very high, possibly close to
the Grand Unification scale. Therefore, the rates for LFV
processes induced by this operator are probably too small to be observed
experimentally \cite{meg_SM}.

On the other hand, there could exist violation of just the
individual family lepton numbers, $\Delta L_i \neq 0$, while
preserving total lepton flavor number, $\Delta L=0$.  This
  violation could be generated at a lower scale than $\Lambda$ in
(\ref{five}). In
this paper we will concentrate on contact four-lepton interactions, that are unrelated to neutrino masses, and that could generate
rates for the LFV processes at observable levels. In fact, this turns
out to be the case in most extensions of the Standard Model, such as
supersymmetry.  We would like to stress that, throughout the paper, by
LFV we mean violation of just the family lepton numbers; total lepton
number is conserved.

We will assume that the physics underlying the effective theory
respects the standard $SU(3)_C \times SU(2)_L \times U(1)_Y$ gauge
symmetry. (Some of the problems that originate when dealing with
non-gauge invariant interactions at low energies have been
discussed in Ref.\cite{DeRujula}.) This simple and reasonable
assumption has a very profound consequence: the same operator
induces several processes, that are related by gauge invariance.
Therefore, one could use the power of gauge invariance to
translate constraints on well studied processes into constraints
on more poorly measured processes, or even not measured at all.
This rationale has been applied in the past to constrain
non-standard neutrino interactions \cite{non-standard-nu}, to rule
out LFV interactions as an explanation for the LSND anomaly
\cite{Bergmann:1998ft} or the atmospheric neutrino anomaly
\cite{Bergmann:1999pk}, and to study the prospects to observe LFV
in a future muon or electron collider \cite{Kabachenko:1997aw}.

In this paper we will try to be more ambitious,
undertaking this analysis for {\it all} the possible
LFV effective interactions, involving only leptons and
compatible with the SM gauge symmetry. 
To this end, we will classify all the possible operators that
can induce LFV processes, and we will list different LFV
processes induced by those operators, together with their
present experimental bounds, if these exist. As explained
above, gauge invariance relates some of these processes and
their corresponding bounds. We will construct tables where
these relations could be used systematically, in order to
translate the bounds on the best constrained LFV processes
into bounds on the worst constrained ones.

The paper is organized as follows.
In the next section we work out the basis of the LFV operators. The tables
are introduced in Section III, where of course we explain how to use
them. In this section we also study the possibility of obtaining
bounds from processes where our operators enter at 
one loop. We devote section IV to some comments and to
present our conclusions.

\section{Operator basis}

We are interested in dimension-six operators involving four lepton
fields with a current$\times$current structure. The restrictions
imposed by gauge invariance on these operators constructed just with 
SM fields was studied by Buchm{\"u}ller and Wyler in
\cite{Buchmuller:1985jz}. According to this reference, 
four classes of operators can be built involving four lepton fields,
namely,  

\begin{eqnarray}
A_{ijkl}&=&
(\bar{L}_i\gamma^\mu L_{j})(\bar{L}_{k}\gamma_\mu L_{l})~,
\label{listA} \\
B_{ijkl}&=&
(\bar{L}_i\gamma^\mu \vec \sigma L_{j})(\bar{L}_{k}\gamma_\mu \vec
\sigma L_{l}) ~,
\label{listB}  \\
C_{ijkl}&=&
(\bar{i}\gamma^\mu j)(\bar{k}\gamma_\mu l)~,
\label{listC} \\
D_{ijkl}&=&
(\bar{i}\gamma^\mu j)(\bar{L}_{k}\gamma_\mu L_{l})~.
\label{listD}
\end{eqnarray}
Here the subindices $i,j,k,l$ refer to the different lepton flavors
$(e,\mu,\tau)$. For one of these flavors $i$, the left-handed lepton doublet
is denoted by $L_i$, and the right-handed singlet by $i$. In
eq.(\ref{listB}), $\vec \sigma=(\sigma^1,\sigma^2,\sigma^3)$ are the
Pauli matrices.

There are indeed other structures that can be formed with four lepton
fields, like
\begin{equation}
(\bar{L}_i \; j)(\bar{k}\; L_{l}) ~.
\label{extraD}
\end{equation}
As discussed in \cite{Buchmuller:1985jz}, one can use the Fierz identity
\begin{equation}
(1 \pm \gamma^5)_{ab}(1 \mp \gamma^5)_{cd}=
\frac{1}{2}[(1 \pm \gamma^5)\gamma^\mu]_{ad}
[(1 \mp \gamma^5)\gamma_\mu]_{cb}~,
\end{equation}
so that eq.(\ref{extraD}) converts trivially to $D_{kjil}$.

However, not all the operators appearing in eqs.(\ref{listA}-\ref{listD})
are independent.
In fact, it turns out that the set of $A$ and the set of
$B$ are linearly dependent. One has to make
use of a Fierz identity for the Pauli $\sigma$ matrices,
\begin{equation}
\vec \sigma_{ab}\vec \sigma_{cd}
=2\delta_{ad}\delta_{cb} - \delta_{ab}\delta_{cd}~,
\end{equation}
and a Fierz identity for the Dirac $\gamma$ matrices,
\begin{equation}
[(1 - \gamma^5)\gamma^\mu]_{ab}[(1 - \gamma^5)\gamma_\mu]_{cd}=- \,
[(1 - \gamma^5)\gamma^\mu]_{ad}[(1 - \gamma^5)\gamma_\mu]_{cb}~.
\end{equation}
With this, it is easy to prove the relation
\begin{equation}
\left(
\begin{array}{c} B_{ijkl}\\B_{ilkj} \end{array} \right) \left( \begin{array}{cc} -1 & 2 \\ 2 & -1 \\ \end{array} \right)
\left( \begin{array}{c} A_{ijkl} \\ A_{ilkj} \end{array} \right)~,
\end{equation}
valid for $j \neq l$. If $j=l$, then $B_{ijkj}=A_{ijkj}$ (no sum
over $j$).

For our analysis, we have to choose one of the two sets; for convenience
we will keep the set $A$ and dispose of the set $B$. Of course, in the
context of a specific data set, it may prove convenient to use $B$
instead of $A$, or even to use all of them keeping in mind the linear
dependence.

In each one of the sets, $A_{ijkl}$, $C_{ijkl}$, and $D_{ijkl}$,
the subindices run over lepton flavor $e,\mu,\tau$.
Some operators are lepton family conserving. Even if these
are beyond the Standard Model too, our aim is to study LFV,
so we concentrate in lepton-family number violating
operators. They may change lepton-flavor numbers by one unit $|\Delta
L_i|=1$ or two units, $|\Delta L_i|=2$ .

In order to proceed and select a basis in the operator space,
we have to take into account that
not all possible combinations of flavor indices lead to independent
operators.
For example, trivially $A_{ijkl}=A_{klij}$. The same happens for
$C$, but not for $D$. In addition, complex conjugation does not
really lead to a new operator, in the sense that
it is obviously the real combination
$g\, O\, +\, g^*\, O^\dagger$
what appears in the Lagrangian.
Finally, in the case of the $C$-type operators,
Fierz rearrangements still lead to relations among different
flavor combinations.

We have carefully taken into account all these arguments and
reached the following conclusion: in the case of $n$ flavors,
there are $n^2(n^2-1)/4$ LFV $A$-type operators,
$n(n^2-1)(n+2)/8$ LFV $C$-type operators and
$n(n-1)(n^2+n-1)/2$ LFV of $D$-type. This makes a total of
$n(n-1)(7n^2+9n-2)/8$ LFV operators.

It is useful to enumerate the independent operators
in the case of two flavors $n=2$, that for the sake of definition
we take as $e$ and $\mu$. We find three $A$-type operators, two
that lead to $|\Delta L_e|= |\Delta L_\mu| =1$, namely
$A_{eee\mu}$ and $A_{e\mu\mu\mu}$, and one leading to
$|\Delta L_e|= |\Delta L_\mu| =2$, which is $A_{e\mu e\mu}$.
In the $C$ sector, we also find a total of 3 operators
that coincide with the order we have shown for $A$.
Finally, there are five independent $D$-operators:
$D_{eee\mu}$, $D_{e\mu ee}$,  $D_{e\mu \mu\mu}$, and
 $D_{\mu\mu e\mu}$ with $|\Delta L_i| =1$ and
$D_{e\mu e\mu}$ with $|\Delta L_i| =2$. Notice our convention
that, among all possible combinations leading to an equivalent
operator, in the first two indices we  
choose the one that has $e$ preceding $\mu$.  
We will follow the convention of putting the first two indices in 
 increasing generation number, namely, $e$ will precede $\mu$ and $\tau$, and
$\mu$ will precede $\tau$. When the two first indices are equal,
  we choose the two last indices in increasing generation number.

Since there are three flavors in nature, there are in total
18 $A$-type operators
(among them 3$\times$2=6 involving actually only two flavors),
15 $C$-type (with also 3$\times$2=6 involving only two flavors),
and 33 $D$-type (with 5$\times$2=10 involving only two flavors).
Therefore, the basis has a total of 66 LFV operators.

The effective Lagrangian describing LFV processes with total lepton
number conserved is
\begin{equation}
{\cal L}=\frac{4G_F}{\sqrt{2}}\sum[a_{ijkl}A_{ijkl}+ c_{ijkl}C_{ijkl} +
d_{ijkl}D_{ijkl} +h.c.\\ ] ~.
\label{efflag}
\end{equation}
We have chosen a convenient normalization that will simplify
future expressions. The adimensional
coefficients $a$, $c$, and $d$ are complex in general.
They contain information about the high-energy scale
$\Lambda$ at which the effective Lagrangian arises,
\begin{equation}
\f{4G_F a_{ijkl}}{\sqrt{2}} \propto \f{1}{\Lambda^2}~,
\end{equation}
and similarly for $c$'s and $d$'s.
\section{LFV processes}

In the previous section we have introduced a minimal basis
of operators that define the most general LFV effective Lagrangian
compatible with the SM gauge symmetry, eqs.(\ref{listA}-\ref{listD}).
For a given process, there are in general several
LFV operators in our basis that induce it. 
Conversely, given one operator, it induces in general
several LFV processes. 
This allows to relate different LFV reactions. Therefore, the experimental 
information available on one of them could be used to constrain
other related reactions. Let us illustrate this with an example.

We consider in detail the exotic $\mu$-decay channel
\begin{equation}
\mu^+\rightarrow e^+ \bar{\nu_e} \nu_e~,
  \label{exomu}
\end{equation}
that was discussed in \cite{LSND}
in the context of the LSND experiment on $\bar{\nu}_e$
appearance in a $\bar{\nu}_\mu$ beam.
Indeed, the LFV decay (\ref{exomu}) could be an alternative
explanation of the anomalous LSND results \cite{Aguilar:2001ty}
without invoking a fourth neutrino. In fact, this is by now past
history since the present bounds on (\ref{exomu}) exclude such an
alternative explanation. Independently of the LSND experiment,
this decay is still of interest to us since it is related
by $SU(2)_L$ gauge invariance to the better measured process
$\mu\rightarrow 3e$, as discussed by
Bergmann and Grossman \cite{Bergmann:1998ft}, and constitutes
a beautiful example of the point that we want to stress in our paper.

Let us discuss now for this particular example
how to derive relations among different
processes using the effective operators introduced in section II.
Clearly, a gauge transformation cannot change family lepton number,
therefore gauge invariance only relates processes with the
same number of leptons and anti-leptons of the same generation.
For instance, the process $\mu^+\rightarrow e^+ \bar{\nu_e} \nu_e$,
that involves three leptons with $|L_e|=1$ and one
lepton with $|L_{\mu}|=1$, is related by gauge invariance
to $\mu^+\rightarrow e^+e^-e^+$ that also involves three leptons with
$|L_e|=1$ and one lepton with $|L_{\mu}|=1$, but not to
$\mu^+\rightarrow e^+ {\bar \nu_e} \nu_{\mu}$, since it involves two
leptons with $|L_e|=1$ and two with $|L_{\mu}|=1$. This classification
is reflected in our tables. 

It can be checked that only the operators $A_{eee\mu}$ and $D_{e\mu ee}$
contribute to the process $\mu^+\rightarrow e^+ \bar{\nu_e} \nu_e$.
Assuming that one of these operators dominates over the other,
the branching ratio for this process is equal to either
$|a_{eee\mu}|^2$ or $|d_{e\mu ee}|^2$, following the normalization
chosen in  eq.(\ref{efflag}). (We will come back to the assumption
of no fine tuned cancellations later on.) Using table I,
it is straightforward to find which other processes
the operators $A_{eee\mu}$ and/or $D_{e\mu ee}$ induce . For instance,
it can be readily checked that $\am\go\aee\eee\aee$ can be induced
by $A_{eee\mu}$ and $D_{e\mu ee}$, and also by
$C_{eee\mu}$ and $D_{eee\mu}$.
The experimental information about these processes, when available,
is shown in the last column of the tables. In this specific case,
there is a stringent experimental bound on $\am\go\aee\eee\aee$
\cite{Eidelman:2004wy} which leads to
\begin{eqnarray}
a^2_{eee\mu} &<& 0.5\e{12}~, \label{croix} \nonumber\\
c^2_{eee\mu} &<& 2\e{12}~,  \nonumber\\
d^2_{e\mu ee}&<& \e{12}~,\nonumber \\
d^2_{eee\mu} &<& \e{12}~.
\end{eqnarray}
For the sake of simplicity in the notation,
we will understand here and in what follows the
modulus squared of a coefficient when just the square is written.
Notice that we are able to constrain each coefficient separately because
of our assumption of barring fine-tuned cancellations. As a result, we
find that there are other LFV processes severely limited by eq.(\ref{croix}).
In particular the limit on the branching ratio of
the decay $\mu^+\rightarrow e^+ \bar{\nu_e} \nu_e$
coming from eq.(\ref{croix}) is
${\cal O}(10^{-12})$, and is much more restrictive than the
direct experimental limit obtained in \cite{Armbruster:2003pq}, which leads to
$a^2_{eee\mu}\leq 9\e{4}$, $d^2_{e\mu ee}\leq 9\e{4}$.
All this information can be read from table I. Furthermore,
from the table we also find that eq.(\ref{croix})
leads to strong restrictions on $\eee\aee\go e^\pm \mu^\mp$,
$\eee\eee\go\eee\m$, etc.
We can use our arguments to bound processes like 
$\nu_e e\go \nu_e \mu$, which are extremely difficult to observe.  

This procedure can be straightforwardly applied to other
processes in the tables. Table I contains processes involving the two
flavors $\mu$ and $e$, and in the next tables we list processes
involving $\tau$ and $e$ (table II), and $\tau$ and $\mu$ (table
III). Finally, table IV is devoted to processes with the three flavors at
work. Each table is subdivided according to the family lepton number
of the particles participating in the process.

Let us summarize the use of the tables. First, one has to check  the
electronic, muonic and tauonic lepton number that are involved in the
process we want to study and look at the corresponding subdivision in
the tables. Once we have found the process we are interested in, we
can read from the table the operators that contribute to that 
process. Barring cancellations, the experimental constraints shown in
the table apply to all the operators separately, and those constraints
on the operators  can be translated into constraints on any other
process in the table.

Notice in passing that for all the subdivisions,
there is at least one process for which experimental information
is available. However, this is not enough to constrain all
the relevant operators. Using just the published bounds on
different LFV processes, and the limits from requiring that
the deviation from the SM prediction
of the decay rates  $\Gamma(\tau^+ \rightarrow e^+ \nu_e
\bar{\nu}_\tau)$ and $\Gamma(\tau^+ \rightarrow \mu^+ \nu_\mu
\bar{\nu}_\tau)$ is smaller than the experimental uncertainty,
it is possible to constrain  all the operators, except one operator of
the type A, seven of the type C and seven of the type D.

\begin{figure}[htbp]
  \centering
  \includegraphics{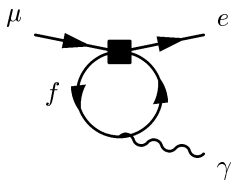}
  \caption{Diagram for $\mu \rightarrow e \gamma$, $(f=e,\mu,\tau)$}
\label{fig1}
\end{figure}

Up to now, we have been considering the effective Lagrangian 
(\ref{efflag}) at tree level. We will see now that some of the operators 
mentioned in the last paragraph can be constrained, 
up to an order of magnitude, using loop-induced processes.
The idea is that in  some of the four-lepton vertexes we can take
two of the external lines, close them to form a loop, and attach
an external gauge boson. (Notice that for operators with $|\Delta L_i|=2$
no loop can be closed.) The resulting process is LFV.
Take for example $\mu\go e\gamma$. It can be generated 
through diagrams like the one shown in Fig. \ref{fig1}.

This transition is absent at tree level and thus
the loop must be finite (we have checked explicitly that
the divergence of the diagram cancels).
Unfortunately, the precise value of the total contribution to the
decay cannot be computed,
due to the presence of unknown counterterms. Nevertheless,
one can estimate an order of magnitude
in terms of the coefficients of the four-lepton
interactions \cite{Raidal:1997hq}. Among  all the possible operators
in our classification, eqs.(\ref{listA}-\ref{listD}),
this transition is induced by $D_{e\mu ee}$, $D_{eee\mu}$, $D_{\mu\mu
  e\mu}$, $D_{e\mu \mu\mu}$, $D_{e\tau\tau\mu}$, and $D_{\mu\tau\tau e}$.
The resulting width is
\begin{eqnarray}
\Gamma (\mu\go e\gamma) \sim \e{4}\alpha\; G_F^2 m^2_f m_\mu^3 |d|^2~,
\end{eqnarray}
with $d$ representing any of the couplings of the relevant operators
and $m_f$ the mass of the fermion running in the loop.
The stringent experimental limit on $\mu\go e\gamma$
\cite{Eidelman:2004wy} allows to put the following order of magnitude upper bounds.

\begin{eqnarray}
d^2_{eee\mu},d^2_{e\mu ee}&<& \e{5} \label{limmegamma}~, \nonumber \\
d^2_{e\mu\mu\mu},d^2_{\mu\mu e\mu}&<&\e{9} \nonumber~, \\
d^2_{\tau \mu e\tau},d^2_{e\tau \tau\mu}&<&\e{11} ~.
\end{eqnarray}

Having $\mu$ and $e$ as external lines, we can also attach a $Z$-boson
instead of a photon and thus we will have the LFV process $Z\go
\mu^\pm e^\mp$. The contribution from the loop gives 
\begin{equation}
  \label{Zem}
\Gamma(Z\go\mu^\mp e^\pm)\simeq 10^{-6}G_F^3M_Z^7|a|^2 ln^2\frac{\Lambda^2}{m^2}\end{equation}
with $a$ standing for the relevant coupling of any of the three types. The fact that $Z$ is massive implies some important differences between
the vertexes $\mu e \gamma$ and $\mu eZ$. While $\mu\go e\gamma$ is a
magnetic dipole transition, for the Z the leading contribution to the
amplitude is of the type 

\begin{equation}\bar{\mu}\gamma^\mu(F_1^V+F_1^A\gamma^5)\, e\, Z_\mu
\label{form1}
\end{equation}

The loop in the effective theory is calculated at $q^2=M_Z^2$ and has
the logarithm appearing in (\ref{Zem}). Also, while for $\mu\go e\gamma$ the
external muon and electron must have different quiralities, this is no longer
the case for (\ref{form1}), so there might be now contributions from the
three types of operators. To be conservative, when finding numerically 
our bounds, we will set
$ln(\Lambda^2/m^2)=ln(\Lambda^2/M_Z^2)$ equal to 1.

The list of contributing operators to $Z\go\mu^\pm e^\mp$ 
is shown in table \ref{tab:5}. The
LEP limits on this exotic vertex \cite{Eidelman:2004wy} 
lead to bounds of
order 10 on the corresponding modulus square coefficients. With the table,
we can compare the limits coming from $\mu\go e\gamma$ to
the corresponding ones from $Z\go\mu e$. When an operator
contributes to both (necessarily of the D-type) the former are much
stronger that the latter {(except in the $D_{e\tau ee}$ and
  $D_{eee\tau}$ cases). However, for the operators contributing to
$Z\go\mu e$ but not to, the limits coming from $Z$ decays are
relevant, although unfortunately not very restrictive.

The same analysis can be done for $\mu-\tau$ and $e-\tau$
transitions, mediated by either $\gamma$ or $Z$. Again, the relevant
information is in table \ref{tab:5}. One can draw conclusions from
the table. For example one can see
that magnetic dipole transitions where a $\tau$ runs in the loop lead
to quite stringent bounds. The physical reason stems from the
necessary helicity-flip in the loop of those transitions. The
interested reader can draw other conclusions from the table. 

Let us stress that the limits in table \ref{tab:5} are just up to an
order of magnitude. Despite this limitation, we find them very useful,
since some operators 
that could not be constrained using tree level processes can be
constrained using loop effects. For instance this is the case for
many operators that involve more than one fermion in the third
generation, such as $A_{e\mu\tau\tau},C_{e\tau\tau\tau},D_{\tau\tau
  e\mu}$, etc.
At the end of the day, combining results from tree level and one
loop processes, we find that among the 66 independent operators 
that contribute
to the Lagrangian, only very few remain unconstrained: three of
type $C$ and two of type $D$. These are $C_{e\tau e\tau}$,
$C_{\mu\tau \mu\tau}$, $C_{e\tau \mu\tau}$, $D_{e \tau e \tau}$,
and $D_{\mu \tau \mu \tau}$. All of them, except $C_{e\tau
\mu\tau}$, could be constrained in the future using $e^- e^-$ and
$\mu^- \mu^-$ linear colliders.
\section{Final comments and conclusions}

In this section, we would like to comment 
on two of our assumptions: the absence of
cancellations and the imposition of gauge invariance. Finally, we
present the conclusions.

When discussing the decay $\mu^+ \rightarrow e^+ \bar{\nu_e} \nu_e$
we did not allow for fine tuned cancellations. Even if unnatural, such
cancellations may spoil some conclusions. In the example we presented
before, the limits in eq.(\ref{croix}), coming from the $\mu \go 3e$
decay, do not hold if there is a destructive interference. For
instance, one could have two operators, $A_{eee\mu}$ and $C_{eee\mu}$, that contribute in such a way that
\begin{equation}
|a_{eee\mu}-\f{1}{2}c_{eee\mu}|^2 \leq \e{12}  ~,
\end{equation}
while $|a_{eee\mu}|^2 \gg \e{12}$ and $|c_{eee\mu}|^2\gg\e{12}$. This
would mean that there would be a contribution to the decay
$\mu^+\rightarrow e^+ \bar{\nu_e} \nu_e$ much larger than
our arguments concluded. We find this possibility very contrived,
although cannot be excluded from first principles.

In the future there might be a positive signal in one (or more) of the
processes listed in the tables. In this case, it will be desirable
to push our type of analysis without the assumption of fine-tuned
cancellations. To do that, we should work with more observables than
just a branching ratio or a cross-section; we should consider for
example polarization measurements. Since these involve different
combinations of operators, one would be able to do a full
statistical analysis and bound all the coefficients
of the relevant operators.

As a final comment, we would like to stress that all the
relations we have presented in the paper are based on
the gauge invariance of the
operators. Although we find this a very reasonable assumption, only
the experiments will tell whether the relevant operators
respect gauge invariance or not. A violation of these relations
would be an indication for extra effects not considered in the
present analysis. For instance, it could happen that for some
reason dimension-six operators were forbidden, and the lowest
dimension operators appear at dimension seven or eight
\cite{non-standard-nu}. In this case, the relations presented in this paper would not hold. 

In conclusion, we have presented a systematic approach to relations
between LFV processes related by the SM gauge symmetry. We have restricted
ourselves to purely leptonic processes. The strategy has been to
build the most general effective Lagrangian that preserves
$SU(2)_L\times U(1)_Y$ and leads to purely leptonic physics. We
have considered operators with energy dimension equal to six,
which can be rearranged to the shape of current$\times$current. We
have introduced no new fields neither right-handed neutrinos. The
special shape of this effective extension has allowed to relate
different LFV processes among themselves, and this has been
applied to the general study of all the 66 operators involving the
three flavors. We have proposed a systematic framework for
studying these relations by means of the use of tables where the
different LFV processes are listed. We have also discussed the
implications coming from loop-induced processes.
In the tables, one can review the current bounds for leptonic LFV searches.


\begin{acknowledgments}
We thank Santi Peris for very useful discussions.
E.M. and J.R. acknowledge support from the CICYT Research Project
FPA2002-00648, from the EU network on Supersymmetry and the Early
Universe (HPRN-CT-2000-00152), and from the \textit{Departament
d'Universitats, Recerca i Societat de la Informaci{\'o}} (DURSI),
Project 2001SGR00188. They are also grateful for hospitality at
CERN where part of this work was carried out.
\end{acknowledgments}


\begin{thebibliography}{99}
\bibitem{rond:2003}
These bounds are obtained from requiring that the 
deviation from the SM predictions of the decay rates
$\tau^+\go\mu^+\bar{\nu}_\mu \nu_\mu$ and  $\tau^+\go
e^+\bar{\nu}_\mu \nu_e$  are smaller than the experimental uncertainties
\cite{Eidelman:2004wy}.
\end{thebibliography}


\newpage

\begin{table}[h]
\begin{tabular}{|r|c||c|c|c||c|}
\hline
& & $A$ & $C$ & $D$ & exp. bound \\
 \hline \hline

$2e+2\mu$& $BR(\mu^- \rightarrow e^- \nu_e \bar{\nu}_\mu$) & $4a_{e\mu e\mu}^2$ & -  & $d^2_{e\mu e\mu}$ &  $<9\times 10^{-4}$\cite{Armbruster:2003pq}\\
\lain
&  $\tilde{P}_{\bar{M}\leftrightarrow M}$ & $ a_{e\mu e\mu}^2 $ &
$c_{e\mu e\mu}^2$ &$ d_{e\mu e\mu}^2$  &  $<3.3 \e{6}$\cite{Willmann:1998gd} \\
\lain
& $\sg{\eee\eee\go\m\m}$ & $12a_{e\mu e\mu}^2$  & $12c_{e\mu e\mu}^2$ & $2d_{e\mu e\mu}^2 $ &  \\
\lain
& $\sg{\nee\eee\go\nm\m}$ & $6a_{e\mu e\mu}^2$ & - & $\f{1}{2}d_{e\mu e\mu}^2$ & \\
\lain
& $\sg{\anm\eee\go\ane\m}$ & $2a_{e\mu e\mu}^2$  & -  &$\f{3}{2}d_{e\mu e\mu}^2$  & $<9\e{3}$\cite{Formaggio:2001jz} \\
\hline \hline
$3e+1\mu$ & $BR(\m\go\eee\nus{e}{e})$ & $a_{eee\mu}^2$ & -  &$d_{eee\mu}^2$ & $<9 \e{4}$\cite{Armbruster:2003pq} \\
\lain
& $BR(\m \go \eee\aee\eee)$ & $2a_{eee\mu}^2$  & $2c_{eee\mu}^2$ & $d_{eee\mu}^2,d_{e\mu ee}^2$ & $<\e{12}$\cite{Bellgardt:1987du} \\
\lain
& $\sg{e^+ e^-\go e^{\pm}\mu^{\mp}}$ & $4a_{eee\mu}^2 $  & $4c_{eee\mu}^2$ & $4d_{eee\mu}^2,4d_{e\mu ee}^2 $ & $<2.3\e{4}$\cite{Akers:1995gz} \\
\lain
& $\sg{\eee\eee\go\eee\m}$ & $6a_{eee\mu}^2 $ & $6c_{eee\mu}^2$ & $2d_{eee\mu}^2,2d_{e\mu ee}^2 $ &  \\
\lain
& $\sg{\nee\eee\go\nee\m}$ & $\f{3}{2}a_{eee\mu}^2$ & - & $\f{1}{2}d_{eee\mu}^2$ & \\
\lain
& $\sg{\ane\eee\go\ane\m}$ & $\f{1}{2}a_{eee\mu}^2$  & -  &$\f{3}{2}d_{eee\mu}^2$  &  \\

\hline \hline
$1e+3\mu$ & $BR(\mu\rightarrow e \nus{\mu}{\mu})$ & $a_{e\mu\mu\mu}^2$ & - & $d_{e\mu\mu\mu}^2$& \\
\lain
 & $\sg{\mu^+\mu^- \go \mu^\pm e^\mp}$ & $4a_{e\mu\mu\mu}^2$ & $4c_{e\mu\mu\mu}^2$ & $4d_{e\mu\mu\mu}^2,4d_{\mu\mu e\mu}^2$ & \\
\lain
& $\sg{\m\m\go\eee\m}$ & $6a_{e\mu\mu\mu}^2$ & $6c_{e\mu\mu\mu}^2$ &
$2d_{e\mu\mu\mu}^2,2d_{\mu\mu e\mu}^2$ & \\
\lain
& $\sg{\nm\eee\go\nm\m}$ & $\f{3}{2}a_{e\mu\mu\mu}^2$ & - & $\f{1}{2}d_{e\mu\mu\mu}^2$ &\\
\lain
& $\sg{\anm\eee\go\anm\m}$ & $\f{1}{2}a_{e\mu\mu\mu}^2$ & - & $\f{3}{2}d_{e\mu\mu\mu}^2$ & $<9\e{3}$\cite{Formaggio:2001jz}\\
\hline

\end{tabular}
\caption{LFV processes involving electron and muon flavors. For the
coefficients, it should be understood modulus squared where
just the square is written (for instance, $a^2_{e\mu e\mu}$ stands
for $|a_{e\mu e\mu}|^2$). $\tilde{P}_{M-\bar{M}}$ represents
the probability of the transition muonium-antimuonium normalized,
for the sake of conciseness in the presentation,
to $P_0=64 e^{12} \left[ m_e^3/G_F m_\mu^5\right]^2=2.56\e{5}$ with
$e$ the electric charge. Therefore, $\tilde{P}_{M-\bar{M}}
=P_{M-\bar{M}}/P_0$. On the other hand, $\tilde{\sigma}(X+Y\go X'+Y')$
is the cross section of the process $X+Y\go X'+Y'$ normalized to
$\sigma_0=G_F^2s/3\pi$, {\it i.e.} $\tilde{\sigma}(X+Y\go
X'+Y')=\sigma(X+Y\go X'+Y')/\sigma_0$. In the tables, near the
threshold for heavy lepton production in electron-(anti)neutrino
scattering, one has to substitute} 
\begin{center}
\renewcommand{\arraystretch}{0.99}
\begin{tabular}{|c||c|c|}
\hline
& $\nu\eee\go\nu l^-$ & $\bar{\nu}\eee\go\bar{\nu} l^-$ \\
\hline\hline
$a^2$ & $a^2(1-\f{m^2_l}{s})^2$ &
$a^2(1-\f{m^2_l}{s})^2(1+\f{m^2_l}{2s})$ \\
\hline
$d^2$ & $d^2(1-\f{m^2_l}{s})^2(1+\f{m^2_l}{2s})$ &
$d^2(1-\f{m^2_l}{s})^2$ \\
\hline
\end{tabular}
\end{center}
\label{tab:1}
\end{table}
\newpage

\pagestyle{empty}
\begin{table}[h]
\begin{tabular}{|r|c||c|c|c||c|}
\hline
& & $A$ & $C$ & $D$ & exp. bound \\
 \hline \hline

$2e+2\tau$ & $\br{\tau^- \go e^- \nus{\tau}{e}}$ & $4a_{e\tau e\tau}^2$ &-  &$d^2_{e\tau e\tau}$ & $<8\e{3}$\cite{rond:2003} \\
\lain
& $\sg{\eee\eee\go\ttt\ttt}$ &$12a_{e\tau e\tau}^2$ & $12c_{e\tau
  e\tau}^2$ &$2d_{e\tau e\tau}^2$ & \\
\lain
& $\sg{\nee \eee \go \nt \ttt} $ & $6a_{e\tau e\tau}^2$ &- &
$\f{1}{2}d_{e\tau e\tau}^2$ & \\
\lain
& $\sg{\ant \eee \go \ane \ttt} $ & $2a_{e\tau e\tau}^2$ &- & $\f{3}{2}d_{e\tau e\tau}^2$ & \\
\hline \hline

$3e+1\tau$ & $\br{\ttt \go \eee \nus{e}{e}}$ & $a_{eee\tau}^2$ & - &$d_{e\tau e e}^2$ &  $<8\e{3}$\cite{rond:2003}\\
\lain
& $\br{\ttt \go \eee\aee\eee}$ & $2a_{eee\tau}^2$  & $2c_{eee\tau}^2$ & $d_{e\tau e  e}^2,d_{eee\tau}^2$ & $<1.6\e{5}$\cite{Bliss:1997iq}  \\
\lain
& $\sg{\eee\aee\go e^\pm \tau^\mp}$ & $4a_{eee\tau}^2 $  & $4c_{eee\tau}^2$ & $4d_{e\tau e e}^2,4d_{eee\tau}^2 $ & $<1.3\e{3}$\cite{Akers:1995gz} \\
\lain
& $\sg{\eee\eee\go\eee\ttt}$ & $6a_{eee\tau}^2 $  & $6c_{eee\tau}^2$ & $2d_{e\tau e e}^2,2d_{eee\tau}^2 $ & \\
\lain
& $\sg{\nee \eee \go \nee \ttt}$ & $\f{3}{2}a_{eee\tau}^2$ &-
&$\f{1}{2}d_{e\tau ee}^2$ & \\
\lain
& $\sg{\ane \eee \go \ane \ttt}$ & $\f{1}{2}a_{eee\tau}^2$ &-
&$\f{3}{2}d_{e\tau ee}^2$ & \\
\lain

\hline \hline

$1e+3\tau$ & $\br{\ttt \go \eee \nus{\tau}{\tau}}$
&$a_{e\tau\tau\tau}^2$ &- &$d_{e\tau\tau\tau}^2$ & $<8\e{3}$\cite{rond:2003} \\
\lain
 & $\sg{\nt \eee \go \nt \ttt}$ &$\f{3}{2}a_{e\tau \tau\tau}^2$ &-
 &$\f{1}{2}d_{e\tau\tau\tau}^2$ & \\
\lain
& $\sg{\ant \eee \go \ant \ttt}$ &$\f{1}{2}a_{e\tau \tau\tau}^2$ &- &$\f{3}{2}d_{e\tau\tau\tau}^2$ & \\
\hline

\end{tabular}
 \caption{LFV processes involving electron and tau flavors.
$\br{\tau\go l l' l''}$ represents the branching ratio of the
corresponding rare tau decay, normalized to $BR(\ttt\go\eee\nus{e}{\tau})$,
{\it i.e.} $\br{\tau\go l l' l''}=BR(\tau\go l l l'')/BR(\ttt\go\eee\nus{e}{\tau})$.
For further explanations in the notation, see table \ref{tab:1} .
}
 \label{tab:3}
\end{table}

\begin{table}[h]
  \centering
\begin{tabular}{|r|c||c|c|c||c|}
\hline
& & $A$ & $C$ & $D$ & exp. bound \\
 \hline \hline
$2\mu+2\tau$ & $\br{\ttt \go \m \nus{\tau}{\mu}}$ &
$4a_{\mu\tau\mu\tau}^2$ &-  &$d^2_{\mu\tau\mu\tau}$ & $<9\e{3}$  \cite{rond:2003}\\
\lain
& $\sg{\m \m \go \ttt \ttt} $ &$12a_{\mu\tau\mu\tau}^2$ &$12c_{\mu\tau\mu\tau}^2$ &$2d_{\mu\tau\mu\tau}^2$ & \\
\hline \hline

$3\mu+1\tau$ & $\br{\ttt \go \m \nus{\mu}{\mu}}$ & $a_{\mu\mu\mu\tau}^2$ & - &$d^2_{\mu\tau\mu\mu}$ & $<9\e{3}$\cite{rond:2003}  \\
\lain
 & $\br{\ttt \go \m \am \m}$ & $2a_{\mu\mu\mu\tau}^2$ & $2c_{\mu\mu\mu\tau}^2$ &$d^2_{\mu\tau\mu\mu},d^2_{\mu\mu\mu\tau}$ & $<1.1\e{5}$\cite{Bliss:1997iq}  \\
\lain
& $\sg{\m \am \go \tau^\pm \mu^\mp}$ & $4a_{\mu\mu\mu\tau}^2$ & $4c_{\mu\mu\mu\tau}^2$ &$4d^2_{\mu\tau\mu\mu},4d^2_{\mu\mu\mu\tau}$ &  \\
\lain
& $\sg{\m\m\go\m\ttt}$ & $6a_{\mu\mu\mu\tau}^2$ & $6c_{\mu\mu\mu\tau}^2$  & $2d^2_{\mu\tau\mu\mu},2d^2_{\mu\mu\mu\tau}$ & \\
\hline \hline

$1\mu+3\tau$ & $\br{\ttt \go \m \nus{\tau}{\tau}}$ & $a_{\mu\tau\tau\tau}^2$ & -  &$d^2_{\mu\tau\tau\tau}$ & $<9\e{3}$\cite{rond:2003}  \\
\hline

\end{tabular}
  \caption{LFV processes involving muon and tau flavors. See explanations
in tables \ref{tab:1} and \ref{tab:3}.}
  \label{tab:2}
\end{table}

\newpage

\thispagestyle{empty}
\renewcommand{\arraystretch}{0.71}
\begin{table}[h]
  \centering
\begin{tabular}{|c|c||c|c|c||c|}
\hline
& & $A$ & $C$ & $D$ & exp. bound \\
 \hline \hline
$2e$  & $BR(\m\go\eee\nus{e}{\tau})$ & $a_{e\mu\tau e}^2$ & - & $d^2_{e\mu\tau e}$ &  \\
\lain
$\Delta L_e=0$ & $\br{\ttt \go \m \nus{e}{e}}$ & $a_{ee\mu\tau}^2$ &  &$d^2_{\mu\tau ee}$ & $<9\e{3}$\cite{rond:2003}  \\
\lain
 & $\br{\ttt \go \eee \nus{e}{\mu}}$ & $a_{e\mu\tau e}^2$ & - &$d^2_{e\tau\mu e}$ & $<8\e{3}$\cite{rond:2003}  \\
\lain
 & $\br{\ttt \go \m\aee\eee }$ & $a_{ee\mu\tau}^2,a_{e\mu\tau e}^2$ & $c_{ee\mu\tau}^2$ & $d^2_{\mu\tau ee},d_{e\tau\mu e}^2,d_{ee\mu\tau}^2,d_{e\mu\tau e}^2$ & $<\e{5}$\cite{Bliss:1997iq} \\
\lain
& $\sg{\eee \aee \go \mu^\pm \tau^\mp} $ & $a_{ee\mu\tau}^2,a_{e\mu\tau e}^2$ & $c_{ee\mu\tau}^2$ & $d^2_{\mu\tau ee},d_{e\tau\mu e}^2,d_{ee\mu\tau}^2,d_{e\mu\tau e}^2$ & $<1.6\e{3}$\cite{Akers:1995gz}  \\
\lain
& $\sg{\nt \eee \go \nee \m}$ &$\f{3}{2}a_{e\mu\tau e}^2$&-& $\f{1}{2}d_{e\mu\tau e}^2$  & \\
\lain
& $\sg{\ane \eee \go \ant \m}$ &$\f{1}{2}a_{e\mu\tau e}^2$&-& $\f{3}{2}d_{e\mu\tau e}^2$  & \\
\lain
& $\sg{\nm \eee \go\nee \ttt}$ & $\f{3}{2}a_{e\mu\tau e}^2$ & -
&$\f{1}{2}d^2_{e\tau\mu e}$ &  \\
\lain
& $\sg{\ane \eee \go\anm \ttt}$ & $\f{1}{2}a_{e\mu\tau e}^2$ & -
&$\f{3}{2}d^2_{e\tau\mu e}$ &  \\

\hline \hline

$2e$ & $BR(\m\go\eee\nus{\tau}{e})$ & $a_{e\mu e\tau}^2$ &- & $d_{e\mu e\tau}^2$ & \\
\lain
$\Delta L_e=2$& $\br{\ttt\go\eee\nus{\mu}{e}}$ & $a_{e\mu e\tau}^2$ &- & $d_{e\tau e\mu}^2$ & $<8\e{3}$ \cite{rond:2003}\\
\lain
& $\br{\ttt\go\eee\am\eee}$ & $2a_{e\mu e\tau}^2$ & $2c_{e\mu e\tau}^2$  & $d_{e\mu e\tau}^2,d_{e\tau e\mu}^2$ & $<9\e{6}$\cite{Bliss:1997iq} \\
\lain
& $\sg{\eee\eee\go\m\ttt} $ & $6a_{e\mu e\tau}^2$ & $6c_{e\mu e\tau}^2$  & $2d_{e\mu e\tau}^2,d_{e\tau e\mu}^2$ & \\
\lain
& $\sg{\nee\eee\go\nt\m} $ & $\f{3}{2}a_{e\mu e\tau}^2$ &-
&$\f{1}{2}d_{e\mu e\tau}^2$ & \\
\lain
& $\sg{\ant\eee\go\ane\m} $ & $\f{1}{2}a_{e\mu e\tau}^2$ &- &$\f{3}{2}d_{e\mu e\tau}^2$ & \\
\lain
& $\sg{\nee\eee\go\nm\ttt} $ & $\f{3}{2}a_{e\mu e\tau}^2$ &-
&$\f{1}{2}d_{e\tau e\mu}^2$ & \\
\lain
& $\sg{\anm\eee\go\ane\ttt} $ & $\f{1}{2}a_{e\mu e\tau}^2$ &- &$\f{3}{2}d_{e\tau e\mu}^2$ & \\
\hline \hline

$2\mu$  & $BR(\m \go \eee \nus{\mu}{\tau})$ & $a_{e \mu\mu\tau}^2$ &-  & $d^2_{e\mu\mu\tau}$ &  \\
\lain
$\Delta L_\mu=0$ & $\br{\ttt \go \m \nus{\mu}{e}}$ & $a_{e\mu\mu\tau}^2$ &-  &$d^2_{\mu\tau e\mu}$ & $<9\e{3}$\cite{rond:2003}  \\
\lain
 & $\br{\ttt \go \eee \nus{\mu}{\mu}}$ & $a_{e\tau \mu\mu}^2$ & - &$d^2_{e\tau \mu\mu}$ & $<8\e{3}$ \cite{rond:2003} \\
\lain
 & $\br{\ttt \go \eee\am\m }$ & $a_{e\tau \mu\mu}^2,a_{e\mu\mu\tau}^2$ & $c_{e\mu\mu\tau}^2$ & $d^2_{e\tau \mu\mu},d_{e\mu\mu\tau}^2,d_{\mu\mu e\tau}^2,d_{\mu\tau e\mu}^2$ & $<\e{5}$\cite{Bliss:1997iq}  \\
\lain
& $\sg{\m \am \go e^\pm \tau^\mp}$ & $a_{e\tau \mu\mu}^2,a_{e\mu\mu\tau}^2$ & $c_{e\mu\mu\tau}^2$ & $d^2_{e\tau \mu\mu},d_{e\mu\mu\tau}^2,d_{\mu\mu e\tau}^2,d_{\mu\tau e\mu}^2$ &  \\
\lain
& $\sg{\nm \eee \go\nt \m}$ &$\f{3}{2}a_{e\mu\mu\tau}^2$ &-
&$\f{1}{2}d_{e\mu\mu\tau}^2$ & \\
\lain
& $\sg{\ant \eee \go\anm \m}$ &$\f{1}{2}a_{e\mu\mu\tau}^2$ &- &$\f{3}{2}d_{e\mu\mu\tau}^2$ & \\
\lain
& $\sg{\nm \eee \go \nm \ttt}$ &$\f{3}{2}a_{e\tau\mu\mu}^2$ &- &$\f{1}{2}d_{e\tau\mu\mu}^2$ & \\
\lain
& $\sg{\anm \eee \go \anm \ttt}$ &$\f{1}{2}a_{e\tau\mu\mu}^2$ &- &$\f{3}{2}d_{e\tau\mu\mu}^2$ & \\
\hline \hline

$2\mu$ & $BR(\m\go\eee\nus{\mu}{\tau})$ & $a_{e\mu\tau\mu}^2$ &- & $d_{e\mu\tau\mu}^2$ & \\
\lain
$\Delta L_\mu=2$& $\br{\ttt\go\m\nus{e}{\mu}}$ & $a_{e\mu \tau\mu}^2$ &- & $d_{\mu\tau \mu e}^2$ & \\
\lain
& $\br{\ttt\go\m\aee\m}$ & $2a_{e\mu\tau\mu}^2$ & $2c_{e\mu\tau\mu}^2$  & $d_{e\mu\tau\mu}^2,d_{\mu\tau\mu e}^2$ & $<9\e{6}$\cite{Bliss:1997iq} \\
\lain
& $\sg{\m\m\go\eee\ttt} $ & $6a_{e\mu\tau\mu}^2$ & $6c_{e\mu\tau\mu}^2$  & $2d_{e\mu\tau\mu}^2,2d_{\mu\tau\mu e}^2$ & \\
\lain
& $\sg{\nt\eee\go\nm\m} $ & $\f{3}{2}a_{e\mu\tau\mu}^2$ &-
&$\f{1}{2}d_{e\mu\tau\mu}^2$ & \\
\lain
& $\sg{\anm\eee\go\ant\m} $ & $\f{1}{2}a_{e\mu\tau\mu}^2$ &- &$\f{3}{2}d_{e\mu\tau\mu}^2$ & $<9\e{3}$\cite{Formaggio:2001jz} \\
\hline \hline

$2\tau$  & $\br{\m \go \eee \nus{\tau}{\tau}}$ & $a_{e\mu\tau\tau}^2$ &-  & $d^2_{e\mu\tau\tau}$ &  \\
\lain
 $\Delta L_\mu=0$& $\br{\ttt \go \m \nus{e}{\tau}}$ & $a_{e\tau\tau\mu}^2$ &-  &$d^2_{\mu\tau \tau e}$ & $<9\e{3}$\cite{rond:2003}  \\
\lain
 & $\br{\ttt \go \eee \nus{\mu}{\tau}}$ & $a_{e\tau\tau\mu}^2$ &-  &$d^2_{e\tau\tau\mu}$ & $<8\e{3}$\cite{rond:2003}  \\
\lain
& $\sg{\nt \eee \go \nt \m}$ &$\f{3}{2}a_{e\mu\tau\tau}^2$ &-
&$\f{1}{2}d_{e\mu\tau\tau}^2$ & \\
\lain
& $\sg{\ant \eee \go \ant \m}$ &$\f{1}{2}a_{e\mu\tau\tau}^2$ &- &$\f{3}{2}d_{e\mu\tau\tau}^2$ & \\
\lain
& $\sg{\nt \eee \go \nm \ttt}$ &$\f{3}{2}a_{e\tau\tau\mu}^2$ &-
&$\f{1}{2}d_{e\tau\tau\mu}^2$ & \\
\lain
& $\sg{\anm \eee \go \ant \ttt}$ &$\f{1}{2}a_{e\tau\tau\mu}^2$ &- &$\f{3}{2}d_{e\tau\tau\mu}^2$ & \\
\hline \hline

$2\tau$ & $\br{\ttt\go\eee\nus{\tau}{\mu}}$ & $a_{e\tau \mu\tau}^2$ &- & $d_{e\tau\mu\tau}^2$ & $<8\e{3}$\cite{rond:2003} \\
\lain
$\Delta L_\tau=2$& $\br{\ttt\go\m\nus{\tau}{e}}$ &
$a_{e\tau\mu\tau}^2$ & - & $d_{\mu\tau e\tau}^2$ & $<9\e{3}$\cite{rond:2003}  \\
\lain
& $\sg{\nm\eee\go\nt\ttt} $ & $\f{3}{2}a_{e\tau\mu\tau}^2$ &-
&$\f{1}{2}d_{e\tau\mu\tau}^2$ & \\
\lain
& $\sg{\ant\eee\go\anm\ttt} $ & $\f{1}{2}a_{e\tau\mu\tau}^2$ &- &$\f{3}{2}d_{e\tau\mu\tau}^2$ & \\
\hline
\end{tabular}
  \caption{LFV processes involving electron, muon and tau flavors..
   See explanations in tables \ref{tab:1} and \ref{tab:3}.}
\label{tab:4}\end{table}

\newpage

\begin{table}[h]
\begin{tabular}{|c|c|c|c|}
\hline
 & e in loop & $\mu$ in loop & $\tau$ in loop\\
\hline
$\mu \go e\gamma$ & $d^2_{eee\mu},d^2_{e\mu ee}<\e{5}$ & $d^2_{e\mu\mu\mu},d^2_{\mu\mu
  e\mu}<\e{9}$ & $d^2_{e\tau\tau\mu},d^2_{\mu\tau \tau e}<\e{11}$\\
\hline
$  \tau \go e\gamma$ & $d^2_{eee\tau},d^2_{e\tau ee}<\E{4} $ &
$d^2_{e\mu\mu\tau},d^2_{\mu\tau e\mu }<\e{1}$ &  $d^2_{e\tau\tau\tau},d^2_{\tau\tau e\tau}<\e{3}$\\
\hline
$  \tau \go\mu\gamma $& $d^2_{e\mu \tau e},d^2_{e \tau\mu e}<\E{4}$ &
$d^2_{\mu\mu\mu\tau},d^2_{\mu\mu \mu\tau}<\e{1}$ &
$d^2_{\mu\tau\tau\tau},d^2_{\tau\tau\mu\tau}<\e{3}$\\ 
\hline
\hline{}
$  Z \go e^\pm\mu^\mp $& $a^2_{eee\mu},c^2_{eee\mu},$ &
$a^2_{e\mu\mu\mu},c^2_{e\mu\mu\mu},$ &
$a^2_{e\tau\tau\mu},a^2_{e\mu\tau\tau},c^2_{e\mu\tau\tau},$\\
 &$d^2_{eee\mu},d^2_{e\mu ee}<10$ & $d^2_{e\mu\mu\mu},d^2_{\mu\mu e\mu}<10$ & $d^2_{\tau\tau e\mu},d^2_{e\mu\tau\tau}<10$ \\
\hline
$  Z \go \tau^\pm e^\mp$ & $a^2_{eee\tau},c^2_{eee\tau},$  & $a^2_{e\mu\mu\tau},a^2_{e\tau\mu\mu},c^2_{e\tau\tau\mu},$ &$a^2_{e\tau\tau\tau},c^2_{e\tau\tau\tau},$ \\
 & $d^2_{eee\tau},d^2_{e\tau ee}<100$ &$d^2_{e\tau\mu\mu},d^2_{\mu\mu e\tau }<100$ & $d^2_{e\tau\tau\tau},d^2_{\tau\tau e\tau}<100$\\
\hline
$  Z \go \mu^\pm\tau^\mp$ & $a^2_{e\mu\tau e },a^2_{ee \mu\tau },c^2_{ee \mu\tau}$ &$a^2_{\mu\mu\mu\tau},c^2_{\mu\mu\mu\tau},$ & $a^2_{\mu\tau\tau\tau},c^2_{\mu\tau\tau\tau},$ \\
&$d^2_{\mu \tau ee},d^2_{ee \mu\tau}<100$ & $d^2_{\mu\mu\mu\tau},d^2_{\mu\mu \mu\tau}<100$ & $d^2_{\mu\tau\tau\tau},d^2_{\tau\tau\mu\tau}<100$\\
\hline
\end{tabular}
\caption{Operators contributing to LFV processes via one-loop
 contributions; the resulting bounds are nothing but an order of
 magnitude.}
\label{tab:5}
\end{table}
\end{document}